\documentclass[letterpaper]{aipproc}
\layoutstyle{6x9}
\usepackage{amsmath,amsfonts,amssymb}
\usepackage{psfrag}
\usepackage{epsfig,floatflt}
\begin{document}
\title{Stability of Homogeneous Extra Dimensions}
\author{Torsten Bringmann}{address={Department of Physics, AlbaNova University Center, Stockholm University, SE - 106 91 Stockholm, Sweden}}
\author{Martin Eriksson}{address={Department of Physics, AlbaNova University Center, Stockholm University, SE - 106 91 Stockholm, Sweden}}
\author{Michael Gustafsson}{address={Department of Physics, AlbaNova University Center, Stockholm University, SE - 106 91 Stockholm, Sweden}}
\begin{abstract}
In order not to be in conflict with observations it is crucial that extra dimensions, if they exist, are stable. It is shown that in the context of homogeneous extra dimensions, this can easily be achieved during both (4D) radiation and vacuum energy dominated eras of the cosmological evolution. During matter domination, however, there is no such possibility even for a very general class of stabilization mechanisms. Even if extra dimensions could be stabilized during matter domination, it is argued that they are generically time-varying during any transition period, such as the one from radiation to matter domination.
\end{abstract}
\maketitle

\newcommand{\dims}{\mbox{($3\!+\!1$)\,}}
\newcommand{\ndims}{\mbox{($3\!+\!n\!+\!1$)\,}}
\newcommand{\be}{\begin{equation}}
\newcommand{\ee}{\end{equation}}

\section{Introduction}

The idea that our world may consist of more than four space-time dimensions, with extra spatial dimensions compactified on some small scale, was first proposed by Nordstr\"om, Kaluza and Klein \cite{nor}. One of the main motivations for these and subsequent works was the hope for a possible unification of all interactions (see e.g.~\cite{bai87} for a review on Kaluza-Klein theories). Recent years have seen a great revival of interest in extra-dimensional scenarios, most notably due to the influence of string theory.

Cosmology provides an important testing ground for theories involving extra dimensions. For example, a generic prediction of these theories is that (some of) the fundamental coupling constants vary with the volume of the internal space. However, the strong cosmological constraints on the allowed variation of these 'constants' (see for example \cite{oli}) require the extra space to be not only compactified, but also stabilized, at a time no later than big bang nucleosynthesis (BBN). One of the main tasks is therefore to find a dynamical explanation for this stabilization and to reproduce standard cosmology for late times.

Here, we will study the prospects for stabilizing homogeneous extra dimensions for a wide range of available stabilization mechanisms. We find that while the extra dimensions can easily be stabilized during both radiation and vacuum energy dominated periods of the cosmological evolution, one encounters generic problems during matter domination. For illustrative reasons, the analysis is done both for the Friedmann-like field equations that are obtained from the higher-dimensional action as well as for the dimensionally reduced and conformally transformed four-dimensional effective action. Of course, these two approaches are equivalent and lead to the same conclusions. The direct comparison, however, helps to clarify hidden subleties and leads to some insight into how one has to interpret the results. For previous work on the stabilization of homogenous extra dimensions see e.g.~\cite{randj,frea} and references therein.

\section{The higher-dimensional Friedmann equations}

In the following we shall adopt the notation and conventions of \cite{bria} and consider a ($3\!+\!1\!+\!n$)\,-\,dimensional spacetime with a metric
\begin{equation}
  g_{AB}~\mathrm{d} X^A\mathrm{d} X^B = - \mathrm{d} t^2 + a^2(t) \gamma_{ij}~\mathrm{d} x^i
    \mathrm{d} x^j + b^2(t) \tilde{\gamma}_{pq}~\mathrm{d} y^p  \mathrm{d}y^q \,,
    \label{metric}
\end{equation}
where $\gamma_{ij}$ and $\tilde{\gamma}_{pq}$ are maximally symmetric
metrics in three and $n$ dimensions, respectively.\footnote{For a separable spacetime the internal space actually \emph{has} to be an Einstein space as a consequence of the field equations \cite{sok}. This slightly more general case does not change our results concerning the stability of the extra dimensions.} Spatial curvature
is thus parametrized in the usual way by $k_a = -1,0,1$ in ordinary
 and $k_b = -1,0,1$ in the compactified space. 
Using such a metric with two time-dependent scale factors $a(t)$ and $b(t)$ is a natural, though obviously rather simplified, way to describe the evolution of the higher-dimensional universe (see, e.g. \cite{bai87}). 

The choice of the metric determines the energy-momentum tensor to be of the form
\be
\label{emtensor} 
  T_{00} = \rho\,, \qquad T_{ij}=-p_a\gamma_{ij}\,,\qquad
  T_{3+p\,3+q}=-p_b\tilde\gamma_{pq}\,,
\end{equation}
which describes a homogeneous but in general anisotropic perfect fluid
in its rest frame. From the field equations,
\begin{equation}
  R_{AB} - \frac{1}{2} R\, g_{AB}+\Lambda\,g_{AB}=
  \kappa^2 T_{AB} \label{fe} \,,
\end{equation}
one now finds the higher-dimensional version of the ordinary Friedmann equations \cite{bria}:

\begin{eqnarray}
\label{allfe}
  3 \left[ \left( \frac{\dot{a}}{a} \right)^2 + \frac{k_a}{a^2} \right] + 3n \frac{\dot{a}}{a} \frac{\dot{b}}{b} +
  \frac{n(n - 1)}{2}\left[\left( \frac{\dot{b}}{b} \right)^2 + \frac{k_b}{b^2} \right] =
  \Lambda + \kappa^2 \rho \,, \label{fe00} \\
   2\frac{\ddot{a}}{a} + \left( \frac{\dot{a}}{a} \right)^2 + \frac{k_a}{a^2} + 2n \frac{\dot{a}}{a}
  \frac{\dot{b}}{b} + n \frac{\ddot{b}}{b} + \frac{n(n - 1)}{2} \left[ \left( \frac{\dot{b}}{b}\right)^2 +
  \frac{k_b}{b^2} \right] = \Lambda - \kappa^2 p_a \,, \label{feij} \\
  \frac{\ddot{b}}{b} + 3\frac{\dot{a}}{a}\frac{\dot{b}}{b} + (n-1)\left[\left(\frac{\dot{b}}{b} \right)^2+\frac{k_b}{b^2}\right] 
  = \frac{2\Lambda}{n+2} + \frac{\kappa^2}{n+2}\left(\rho-3p_a+2p_b\right)\,,
  \label{fe3}
\end{eqnarray}
where a dot denotes differentiation with respect to $t$.

\section{Static solutions and standard cosmology}

One usually assumes that compactification and stabilization of the internal space can be attained dynamically by introducing background fields. Since their role is to separate ordinary space from the extra dimensions they typically contribute an effective action
\be
  \label{sbg}
  S^{\mathrm{bg}}=-\int\mathrm{d}^{4+n}X\sqrt{-g}\,W(b)
\ee
to the theory \cite{bai84a,bai87}. Examples of concrete mechanisms include gauge-fields wrapped around two extra dimensions \cite{cre} and the generalization of this to the Freund-Rubin mechanism \cite{freb}, as well as stabilization by the Casimir effect \cite{appb}.

The energy-momentum tensor corresponding to the action (\ref{sbg}) takes the form
\be
\label{tbg} 
  T^{\mathrm{bg}}_{00} = \rho^{\mathrm{bg}}\,, 
  \qquad T^{\mathrm{bg}}_{ij}=-p^{\mathrm{bg}}_a\gamma_{ij}\,,
  \qquad T^{\mathrm{bg}}_{3+p\,3+q}=-p^{\mathrm{bg}}_b\tilde\gamma_{pq}\,.
\ee
%
Let us now assume that a stabilization mechanism due to background fields
is at work, i.e.~$b(t)=b_0$ is constant at late times. Since these
fields also contribute to the energy-momentum tensor, we replace
$\rho\rightarrow\rho+\rho^{\mathrm{bg}}$ and $p_{a,b}\rightarrow
p_{a,b}+p_{a,b}^{\mathrm{bg}}$ in the field
equations (\ref{fe00}\,-\,\ref{fe3}). All contributions to the energy-momentum
tensor that are not due to background fields or a cosmological
constant are then given by $\rho$ and $p_{a,b}$, and
equations (\ref{fe00}\,-\,\ref{feij}) reduce to the standard
Friedmann equations with an effective four-dimensional cosmological
constant \cite{randj,bai87}
\be
  \label{lambda4}
  \bar\Lambda = \frac{2\Lambda}{n+2} + \frac{\kappa^2}{n+2}\,
     \Big[2W(b_0)+b_0W'(b_0)\Big].
\ee
The remaining equation (\ref{fe3}),
\be
  \label{f3const}
  (n+2)(n-1)\frac{k_b}{b_0^2} = 2\Lambda + 
  2\kappa^2\left[W(b_0)-\frac{b_0}{n}W'(b_0)\right]+
  \kappa^2\left(\rho-3p_a+2p_b\right)\,,
\ee
however, can only be satisfied if the last term is constant \cite{bria}:
\be
  \label{eof}
  \rho - 3p_a + 2p_b = const.
\ee
One possibility to achieve this is during usual four-dimensional radiation domination ($\rho=1/3p_a$, $p_b=0$), which is expected to occur more or less directly after compactification and stabilization has taken place, and $a\gg1/T\gg b$. Another possibility is given by an energy-momentum tensor dominated by vacuum energy (which has constant energy density and pressure).

In all other cases, though, it is hard to see how the constraint (\ref{eof}) can possibly be satisfied. In particular, throughout most of its evolution the universe has been dominated by non-relativistic matter with negligible pressure $p_a\ll\rho$. The only way to get static extra dimensions during such a matter-dominated epoch would be to have 
\be
  p_b=-\frac{1}{2}\rho+const.
\ee
Since $\rho$ is the \emph{total} energy density, such an equation of state seems highly contrived.

\begin{figure}
  \begin{minipage}[t]{0.49\textwidth}
      \centering
      \psfrag{a}[][][0.9]{$a$}
      \psfrag{b}[][][0.9]{$b$}
      \psfrag{r}[][][0.9]{$\frac{\rho_m}{\rho}$}
      \psfrag{x}[t][][0.9]{$\log t$}
      \psfrag{y1}[][][0.9]{$\log a,~\log b$}
      \psfrag{y2}[][][0.9]{$\rho_m/\rho$}
      \psfrag{0}[][][0.75]{$0$}
      \psfrag{2}[][][0.75]{$2$}
      \psfrag{4}[][][0.75]{$4$}
      \psfrag{6}[][][0.75]{$6$}
      \psfrag{8}[][][0.75]{$8$}
      \psfrag{5}[][][0.75]{$5$}
      \psfrag{10}[][][0.75]{$10$}
      \psfrag{15}[][][0.75]{$15$}
      \psfrag{20}[][][0.75]{$20$}
      \psfrag{0.}[][r][0.75]{$0.0$}
      \psfrag{0.1}[][][0.75]{$0.1$}
      \psfrag{0.2}[][][0.75]{$0.2$}
      \psfrag{0.3}[][][0.75]{$0.3$}
      \psfrag{0.4}[][][0.75]{$0.4$}
      \psfrag{0.5}[][][0.75]{$0.5$}
      \psfrag{0.6}[][][0.75]{$0.6$}
      \psfrag{0.7}[][][0.75]{$0.7$}
      \includegraphics[width=\textwidth]{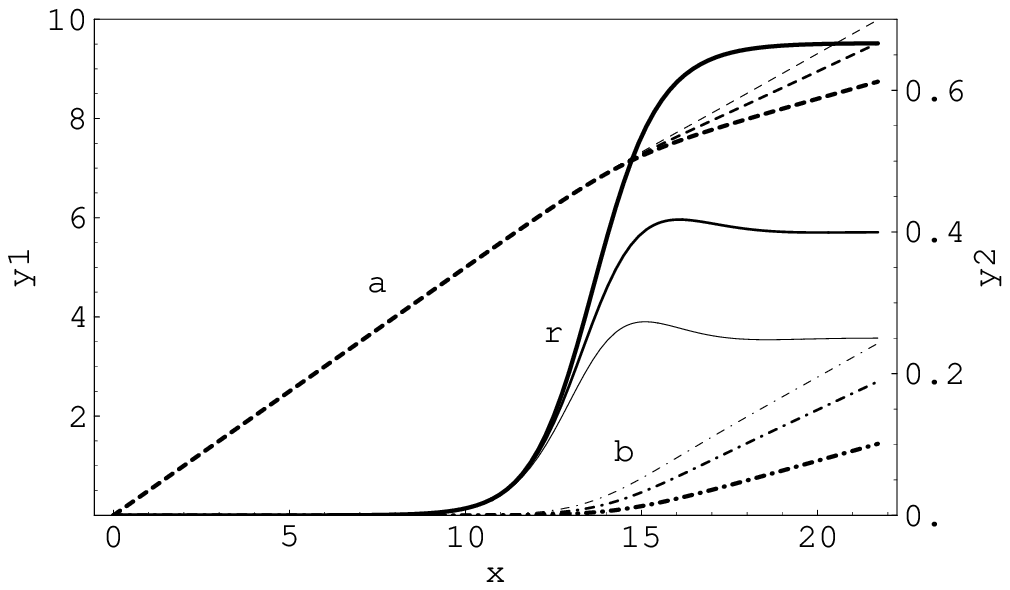}
   \end{minipage}
  \begin{minipage}[t]{0.02\textwidth}
   \end{minipage}
  \begin{minipage}[t]{0.49\textwidth}
      \centering   
       \psfrag{a}[][][0.9]{$a$}
       \psfrag{b}[][][0.9]{$b$}
       \psfrag{r}[][][0.9]{$\frac{\rho_m}{\rho}$}
       \psfrag{x}[t][][0.9]{$\log t$}
       \psfrag{y1}[][][0.9]{$\log a,~\log b$}
       \psfrag{y2}[][][0.9]{$\rho_m/\rho$}
       \psfrag{0}[][][0.75]{$0$}
       \psfrag{2}[][][0.75]{$2$}
       \psfrag{4}[][][0.75]{$4$}
       \psfrag{6}[][][0.75]{$6$}
       \psfrag{8}[][][0.75]{$8$}
       \psfrag{5}[][][0.75]{$5$}
       \psfrag{10}[][][0.75]{$10$}
       \psfrag{15}[][][0.75]{$15$}
       \psfrag{20}[][][0.75]{$20$}
       \psfrag{0.}[][][0.75]{$~~0.0$}
       \psfrag{0.1}[][][0.75]{$0.1$}
       \psfrag{0.2}[][][0.75]{$0.2$}
       \psfrag{0.3}[][][0.75]{$0.3$}
       \psfrag{0.4}[][][0.75]{$0.4$}
       \psfrag{0.5}[][][0.75]{$0.5$}
       \psfrag{0.6}[][][0.75]{$0.6$}
       \psfrag{0.7}[][][0.75]{$0.7$}
       \psfrag{0.8}[][][0.75]{$0.8$}
       \psfrag{0.9}[][][0.75]{$0.9$}
       \psfrag{1.}[][][0.75]{$~~1.0$}
      \includegraphics[width=\textwidth]{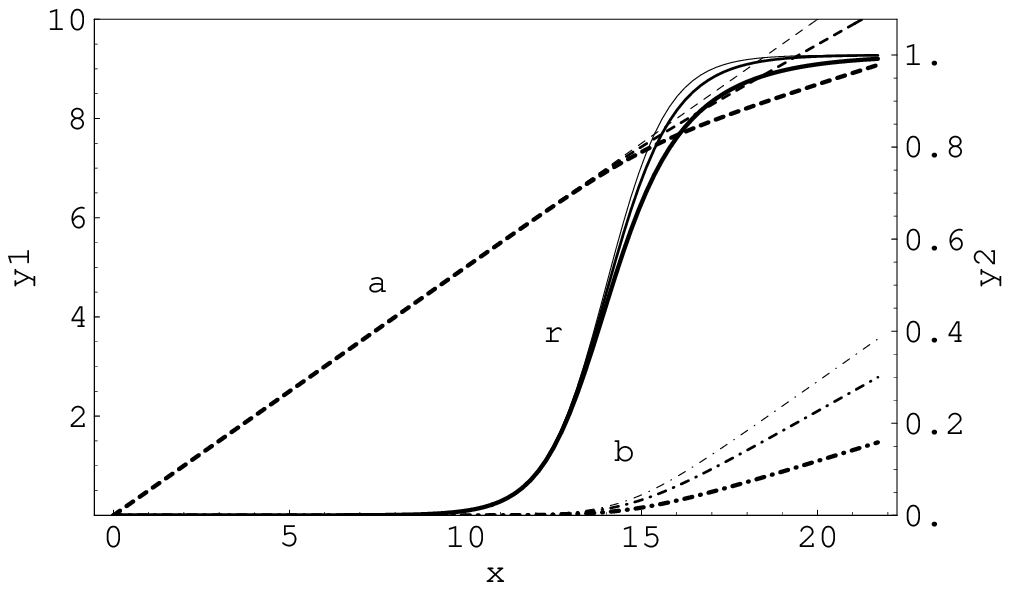}
   \end{minipage}
     \caption{The evolution of the scale
    factors $a$ and $b$, as well as the fractional energy density in non-relativistic matter $\rho_m/\rho$ for $n =$1 (thin), 2
    (medium) and 7 (thick) with $\left(\rho_m/\rho\right)_i = 10^{-7}$.
    The figure on the left is taken from \cite{brib} and shows the case of Kaluza-Klein-like dark matter in a model of so-called universal extra dimensions \cite{app}, where the matter-dominated period is described by an equation of state $p_a=0$, $p_b=1/n$. The right figure shows the case where the KK-contribution to the dark matter is negligible and one has $p_a=p_b=0$ during matter-domination.}
     \label{fig1}
\end{figure}

Having excluded the possibility of exactly static extra dimensions during matter domination, the natural question arises how much the scalefactor $b$ actually is expected to vary and whether this would influence the evolution of the scalefactor $a$ in an unwanted way. Of course, the answer to this will be model-dependent. Figure \ref{fig1} shows the transition from a radiation to a matter-dominated period for two cases of particular interest. For both one finds a considerable variation of $b$ as soon as the non-relativistic matter content makes up more than about 10 percent of the total energy. Note that the evolution of $a$ does not follow the expected pattern either (but see the discussion in the next section).

\section{Dimensional reduction}

Let us start from the higher-dimensional action
\be
  \label{action1}
  S=\frac{1}{2\kappa^2}\int\mathrm{d}^{4+n}X\sqrt{-g}
    \left(R - 2\Lambda - 2\kappa^2\mathcal{L}_\mathrm{matter}\right)
\ee
and assume a metric of a slightly more general kind than (\ref{metric}):
\be
 g_{AB}~\mathrm{d} X^A \mathrm{d} X^B = \bar g_{\mu\nu}(x^\rho)~\mathrm{d} x^\mu \mathrm{d} x^\nu
    +b^2(x^\mu)\tilde g_{pq}(y^p)~\mathrm{d} y^p \mathrm{d} y^q \,. 
\ee
Integrating out the extra dimensions then gives
\be
  \label{action2}
   S=\frac{1}{2\hat\kappa^2}\int\mathrm{d}^{4}x\sqrt{-\bar g}~b^n
    \left(\bar R +b^{-2}\tilde R + n(n-1)b^{-2}\partial_\mu b \partial^\mu b 
    - 2\Lambda - 2\kappa^2\mathcal{L}_\mathrm{matter}\right)\,,
\ee
where $\hat\kappa^2\equiv\kappa^2/\int\mathrm{d}^{n}y\sqrt{\tilde g}$ and $\bar R$ ($\tilde R$) is the curvature scalar constructed from $\bar g_{\mu\nu}$ ($\tilde g_{pq}$). 

The next step is to change to the Einstein frame by a conformal transformation to a new metric $\hat g_{\mu\nu}=b^n \bar g_{\mu\nu}$. This leads to four-dimensional gravity minimally coupled to a scalar field $\Phi\equiv\sqrt{\frac{n(n+2)}{2\hat\kappa^2} }\ln{b}$:
\be
  \label{action3}
  S=\int\mathrm{d}^{4}x\sqrt{-\hat g}
    \left(\frac{1}{2\hat\kappa^2} \hat R 
    - \frac{1}{2}\partial_\mu\Phi \partial^\mu\Phi + V_{\mathrm{eff}}(\Phi)\right)\,,
\ee
where the potential of the scalar field is given by
\be
  \label{veff}
  V_{\mathrm{eff}}(\Phi)=\frac{n(n-1)k_b}{2\hat\kappa^2}~
        \mathrm{e}^{-\sqrt\frac{2(n+2)\hat\kappa^2}{n}\Phi}
        -\frac{1}{\hat\kappa^2}\left(\Lambda+\kappa^2\mathcal{L}_\mathrm{matter}\right)
        \mathrm{e}^{-\sqrt\frac{2n\hat\kappa^2}{n+2}\Phi}\,.
\ee

The theories described by (\ref{action1}) and (\ref{action3}) are
completely equivalent, i.e.~they result in the same equations of
motion. For the scalar field in (\ref{action3}), for example, one has
\be
  \label{box}
  \hat\Box\Phi=-V_{\mathrm{eff}}'(\Phi)\,,
\ee
whith $\hat\Box=\frac{1}{\sqrt{-\hat g}}\partial_ \mu\sqrt{-\hat
  g}\partial^\mu$ being the Laplace-Beltrami operator for the metric
$\hat g_{\mu\nu}$. In order to compare this with the results of the
previous sections, we restrict ourselves to the metric (\ref{metric}) and write the higher-dimensional matter part in perfect fluid form $\mathcal{L}_\mathrm{matter}=\rho=\rho_0a^{-3(1+w_a)}b^{-n(1+w_b)}$. One may now verify that (\ref{box}) gives exactly (\ref{fe3}),
from which all previous results follow \cite{bria}. The equation of motion (\ref{box}) for the scalar field thus provides a convenient way to look for static solutions of the scale factor $b$: since the spatial derivatives vanish, such solutions must obey $V_\mathrm{eff}'(\Phi)=0$, and if the extremal point of $V_\mathrm{eff}(\Phi)$ is a minimum, one even expects these solutions to be stable. 

In principle there are infinitely many, mathematically equivalent frames that are related to each other by a conformal transformation. However, caution must be taken as to which of the conformally related metrics one identifies as the physical one. The requirement that the conformally transformed system in four dimensions has positive definite energy singles out a unique conformal factor \cite{sok,cho}, which is exactly the one that transforms to the Einstein frame (\ref{action3}). This suggests that $\hat g_{\mu\nu}$ should be regarded as the physical metric\footnote{Note that this choice results in the non-minimal coupling of the scalar field to the rest of the matter part, $\mathcal{L}_\mathrm{matter}$, that can be seen in (\ref{veff}) and which is absent in the Jordan frame (\ref{action2}). Sometimes this has been used as an argument why one should choose the Jordan frame as the physical one, see e.g. the discussion given in \cite{cho}. For the explicit view that all conformally related frames are not only mathematically but also physically equivalent see \cite{fla}.}, with $\mathrm{d}\hat t\equiv b^\frac{n}{2}~\mathrm{d}t$ being the measured cosmological time and $\hat a\equiv b^\frac{n}{2}a$ the ordinary 4D scalefactor. As long as one expects approximately static extra dimensions, however, the difference between $a$ and $\hat a$ (or $t$ and $\hat t$) should of course be negligible for all practical purposes.

Another important point to notice here is that $V_\mathrm{eff}(\Phi)$ is time-dependent through its dependence on $\mathcal{L}_\mathrm{matter}$. So even if there existed a minimum for the effective potential at all times of the cosmological evolution, its value would still be expected to vary during transition periods such as from radiation to matter domination. The scale factor $b$ would then no longer stay constant but  be driven towards the new minimum according to (\ref{box}).

\section{Conclusions}
\label{conc}

 It is crucial for the viability of extra-dimensional scenarios that, for late times, the extra dimensions are stabilized and standard cosmology can be recovered. A wide range of possible stabilization mechanisms can be described in a phenomenological way by adding a 'radion potential' of the form (\ref{sbg}) to the action. We have shown that the stabilization of homogeneous extra dimensions in this set-up can easily be achieved for the radiation dominated period of the cosmological evolution as well as during vacuum-energy domination. Surprisingly enough, however, there are generic problems during matter domination. By allowing for different types of stabilization mechanisms, for example of the form $W(b,\rho)$, one might in principle get static solutions \cite{bria} - though certainly at the cost of some serious fine-tuning (remember that is not enough to fix the size of the extra dimensions, one also has to reproduce the correct behaviour for the 4D scalefactor). As discussed in the last section, the size of the extra dimensions would in any case be expected to vary during transition periods.

Problems with the stabilization of extra dimensions even arise in certain brane world models once one allows for non-relativistic contributions to the matter content \cite{arn}. It would be interesting to further investigate the various existing extra-dimensional models and see if and how they can avoid these problems, i.e.~whether they produce the correct late-time cosmology not only for radiation- but also for matter-domination.

\begin{theacknowledgments}
We would like to thank David Wiltshire for helpful discussions on conformal transformations and the organizers of Phi in the Sky for providing a nice and enjoyable atmosphere in Porto.
\end{theacknowledgments}



\begin{thebibliography}{99}

\bibitem{nor}
G. Nordstr\"om, Phys.~Z.~{\bf 15}, 504 (1914);
T. Kaluza, Sitzungsber.~Preuss.~Akad.~Wiss., Phys. Math. Kl., 966 (1921);
O. Klein, Z.~Phys.~{\bf 37}, 895 (1926).

\bibitem{bai87}
D.~Bailin and A.~Love,
Rept.\ Prog.\ Phys.\  {\bf 50}, 1087 (1987).

\bibitem{oli}
W.~J.~Marciano,
Phys.\ Rev.\ Lett.\  {\bf 52}, 489 (1984);
K.~A.~Olive \textit{et al.},
Phys.\ Rev.\ D {\bf 66}, 056022 (2002);
J.~P.~Uzan,
Rev.\ Mod.\ Phys.\  {\bf 75}, 403 (2003)
J.~M.~Cline and J.~Vinet,
Phys.\ Rev.\ D {\bf 68}, 025015 (2003);
C.~J.~A.~Martins \textit{et al.},
Phys.\ Lett.\ B {\bf 585}, 29 (2004).

\bibitem{randj}
S. Randjbar-Daemi, A. Salam and J. Strathdee,
Phys.~Lett.~B {\bf 135}, 388 (1984).
  
\bibitem{frea}
P. G. O. Freund, Nucl.~Phys.~B {\bf 209} 146 (1982);
K. Maeda, Class.~Quant.~Grav.~{\bf 3} 233 (1986);
Maeda K, Class.~Quant.~Grav.~{\bf 3} 651 (1986);
U. Gunther and A. Zhuk, Class.~Quant.~Grav.~{\bf 15}, 2025 (1988);
U. Gunther and A. Zhuk, Phys.~Rev.~D {\bf 61}, 124001 (2000);
S. M. Carroll, J. Geddes, M. B. Hoffman and R. M. Wald,
Phys.~Rev.~D {\bf {66}}, 024036 (2002);
U.~Gunther, A.~Starobinsky and A.~Zhuk,
Phys.\ Rev.\ D {\bf 69}, 044002 (2004).

\bibitem{bria}
T.~Bringmann and M.~Eriksson,
JCAP {\bf 0310}, 006 (2003).

\bibitem{sok}
L.~M.~Soko\l owski,
Class.\ Quant.\ Grav.\  {\bf 6}, 59, (1989).

\bibitem{bai84a}
D. Bailin, A. Love and C. E. Vayonakis,
Phys.~Lett.~B {\bf 142}, 344 (1984).

\bibitem{cre}
E. Cremmer and J. Scherk,
Nucl.~Phys.~B {\bf 108}, 409 (1976);
S. Randjbar-Daemi, A. Salam and J. Strathdee,
Nucl.~Phys.~B {\bf 214}, 491 (1983);
A. Salam and E. Sezgin,
Phys.~Lett.~B {\bf 147}, 47 (1984);
Y. Okada,
Phys.~Lett.~B {\bf 150}, 103 (1985);
S. Randjbar-Daemi, A. Salam, E. Sezgin and J. Strathdee,
Phys.~Lett.~B {\bf 151}, 351 (1985).

\bibitem{freb}
P. G. O. Freund and M. A. Rubin,
Phys.~Lett.~B {\bf 97}, 233 (1980).

\bibitem{appb}
T. Appelquist and A. Chodos,
Phys.~Rev.~Lett.~{\bf 50}, 141 (1983);
T. Appelquist and A. Chodos,
Phys.~Rev.~D {\bf 28}, 772 (1983);
P. Candelas and S. Weinberg,
Nucl.~Phys.~B {\bf 237}, 397 (1984);
D. Bailin and A. Love,
Phys.~Lett.~B {\bf 137}, 348 (1984);
E. Ponton and E. Poppitz, JHEP {\bf 06}, 019 (2001).

\bibitem{brib}
T.~Bringmann, M.~Eriksson and M.~Gustafsson,
Phys.\ Rev.\ D {\bf 68}, 063516 (2003).

\bibitem{app}
T. Appelquist, H. Chang and B. A. Dobrescu,
Phys.~Rev.~D {\bf 64}, 035002 (2001);
H. C. Cheng, K. T. Matchev and M. Schmaltz,
Phys.~Rev.~D {\bf 66}, 036005 (2002);
G. Servant and T. M. Tait,
Nucl.~Phys.~B {\bf 650}, 391 (2003).

\bibitem{cho}
Y.~M.~Cho,
Phys.\ Rev.\ Lett.\  {\bf 68}, 3133 (1992);
V.~Faraoni, E.~Gunzig and P.~Nardone,
Fund.\ Cosmic Phys.\  {\bf 20}, 121 (1999).

\bibitem{fla}
E.~E.~Flanagan,
Class.\ Quant.\ Grav.\  {\bf 21}, 3817 (2004).
[arXiv:gr-qc/0403063].

\bibitem{arn}
R.~Arnowitt, J.~Dent and B.~Dutta,
arXiv:hep-th/0405050.


\end{thebibliography}
\end{document}